\documentclass[aip,reprint]{revtex4-1}
\usepackage{tikz}
\newcommand*\circled[1]{\tikz[baseline=(char.base)]{
            \node[shape=circle,draw,inner sep=1pt] (char) {#1};}}

\usepackage{graphicx,here}
\usepackage{dcolumn}
\usepackage{bm}
\usepackage{mathtools}
\usepackage{color}
\usepackage{ulem}
\usepackage[utf8]{inputenc}
\usepackage[T1]{fontenc}
\usepackage{mathptmx}
\begin{document}
\title{Controlling  the nucleation and annihilation of skyrmions with magnetostatic interactions}
\author{N. Vidal-Silva}\email{nvidalsilva@ing.uchile.cl}
\affiliation{Center for the Development of Nanoscience and Nanotechnology (CEDENNA), 917-0124 Santiago, Chile}%
\affiliation{Departamento de F\'isica, Facultad de Ciencias F\'isicas y Matem\'aticas, Universidad de Chile, Casilla 487-3, Santiago, Chile.}
\author{A. Riveros}
\affiliation{Departamento de F\'isica, Universidad de Santiago de Chile (USACH), Avda. Ecuador 3493, 917-0124 Santiago, Chile}
\author{F. Tejo}
\affiliation{Center for the Development of Nanoscience and Nanotechnology (CEDENNA), 917-0124 Santiago, Chile}%
\affiliation{Departamento de F\'isica, Universidad de Santiago de Chile (USACH), Avda. Ecuador 3493, 917-0124 Santiago, Chile}
\author{J. Escrig}
\affiliation{Center for the Development of Nanoscience and Nanotechnology (CEDENNA), 917-0124 Santiago, Chile}%
\affiliation{Departamento de F\'isica, Universidad de Santiago de Chile (USACH), Avda. Ecuador 3493, 917-0124 Santiago, Chile}
\author{D. Altbir}
\affiliation{Center for the Development of Nanoscience and Nanotechnology (CEDENNA), 917-0124 Santiago, Chile}%
\affiliation{Departamento de F\'isica, Universidad de Santiago de Chile (USACH), Avda. Ecuador 3493, 917-0124 Santiago, Chile}



\date{\today}

\begin{abstract}
Skyrmions have become one of the most visited topics during the last decade in condensed matter physics. In this work, and by means of analytical calculations and micromagnetic simulations, we explore the effect of the magnetostatic field generated by a magnetic tip on the stability of skyrmions. Our results show that the interaction energy between the tip and the skyrmion plays a fundamental role in the stabilization of N\'eel skyrmions confined in nanodisks, allowing its nucleation and annihilation, and also providing a precise control of its size and polarity. Based on our results, we propose a very simple and cyclic method to nucleate and annihilate skyrmions, as well as to control their polarity and chirality. This proposal could open new possibilities for logic devices taking advantage of all the degrees of freedom that skyrmionic textures have.
\end{abstract}	


\maketitle

Magnetic skyrmions are non-uniform spin configurations topologically protected. This important feature has made them proposed for the implementation of soliton-like based race-track memories\cite{NatNanoF13,zhang2015,tomasello2015}. The low current density required to move skyrmions \cite{NatNanoS13,NatNanoF13,tomasello2015,Iwasaki2013,Yu2012}, the low dissipation when their dynamic is activated, and their ability to overcome defects or pinning sites in a sample are very convenient for such technological developments\cite{NatNanoF13,NatNanoS13,SciRep15,SciRep16,NJPhys17}. There are two types of experimentally observed skyrmions \cite{JMMM94,NatMater15}, namely, Bloch (vortex-like) and N\'eel (hedgehog-like) skyrmion. Both  share the general structure of a skyrmion but they differ not only in their shape, but also on how they are usually stabilized: while Bloch skyrmions appear when the chiral Dzyaloshinskii-Moriya interaction (DMI) or the long-range dipolar interaction is present\cite{DA,riveros2018}, N\'eel skyrmions need the existence of an interfacial DMI to be stabilized. Skyrmions also exhibit two important geometric features: their chirality ($C$), which defines the direction in which the spins are oriented from the center towards the edge of the magnetic texture, and their polarity ($P$), which defines the magnetization direction in the core of the magnetic texture.  In principle, the polarity of skyrmions can be controlled through external magnetic fields\cite{riveros2018}, while the chirality depends on the DMI sign, but which in turn is associated with the skyrmion polarity due to the minimization process of the DMI energy. Thus, a positive chirality $C = 1$ (outward) corresponds to a positive polarity $P = 1$ (up), while a negative chirality $C = -1$ (inward) is determined by a negative polarity $P = -1$ (down).



Isolated magnetic skyrmions hosted in cylindrical nanostructures have recently gained a lot of attention due to the benefits that this geometry offers to stabilize these magnetic textures \cite{juge2018,vidal2017,riveros2018,castro2016,guslienko2018,rohart2013,tejo2018,boulle2016}. In the search for spintronics applications, this geometry seems to be a natural candidate to host skyrmions mainly because edge effects, which yield the possibility of stabilizing both types of skyrmions\cite{vidal2017,riveros2018}.

For skyrmions-based future spintronic applications in magnetic nanodots it is fundamental to achieve a perfect control of all degrees of freedom, polarity and chirality. In this sense, a perfect control of the process of writing (nucleation) and deleting (annihilation) of isolated skyrmions could be an ideal scenario for such control. Different methods have been proposed to achieve this goal\cite{Iwasaki2013, NatNanoS13, rooming2013,legrand2017,jiang2015,zhou2014,hsu2017}. For instance, recently Zhang \textit{et al.}\cite{zhang2018} showed that it is possible to create skyrmion lattices by means of Magnetic Force Microscopy (MFM) using the stray field that the MFM tip generates. On the other hand, the stabilization of isolated and confined skyrmions has been achieved by means of geometrical confinement\cite{guslienko2018,castro2016,vidal2017,rohart2013} or by applying external magnetic fields\cite{tejo2018,riveros2018}. 

Recently, Garanin \textit{et al.} \cite{garanin18} studied the nucleation of skyrmions in thin films by means of a dipolar field generated by a nanosized magnetic dipole. In this paper we go further and propose a skyrmion-based data storage device which incorporates not only the nucleation of isolated skyrmions, but also their annihilation and the control of their size, polarity and chirality through the magnetostatic field generated by a ferromagnetic cylindrical tip. Our proposal is based on the fact that an inhomogeneous magnetostatic field in the vicinity of a skyrmion hosted in a nanodot produces radical consequences on its size and can even induce magnetic phase transitions depending on the distance between the skyrmion and the magnetic tip.

Our model considers two cylindrical nanoparticles with saturation $M_{s_i}$, arbitrary thickness $L_i$ and radii $R_i$, where $i = 1,2$ label the dots, separated by a distance $s = \sqrt {d^2 + h^2}$, being $d$ the horizontal distance  measured from their centers and $h$ the vertical separation measured from base to base. We use the micromagnetism theory \cite{aharoni2000}, in which the discrete nature of the magnetic moments is replaced by a continuous field called the magnetization $\mathbf{M}(\mathbf{r})$ and a low temperature regime is assumed. In this framework, and using cylindrical coordinates for the normalized magnetization $\mathbf{m}_i(\mathbf{r}_i) = m_{r,i}(\mathbf{r}_i)\hat{\mathbf{r}} + m_{z,i}(\mathbf{r}_i)\hat{\mathbf{z}}$ for each dot, we obtain a general expression for the magnetostatic interaction energy between two dots separated by an arbitrary distance,  This expression is presented in  Appendix \ref{E_int_ap}. However, for the purpose of this work, we will assume that  the horizontal distance $d = 0$. From now on, the upper dot labelled $``2"$ will be denoted as \textit{the magnetic tip}, and is assumed uniformly magnetized along its symmetry axis, that is the $\pm\hat{z}$-direction. We will also impose that the dot located below the magnetic tip, hereafter denoted \textit{the magnetic dot}, can has the general azimuthal magnetization described above.

Under these assumptions, the magnetostatic interaction energy between the tip and the dot given by Eq. \ref{Eint_ap} takes the form
\begin{eqnarray}
\label{Eint_main}
E_{int} = \pi \mu_0M_{s_1}M_{s_2} \int_0^{\infty}dk g_{0,2}^z(k)\beta(k),\hspace{0.5cm}
\end{eqnarray}
where the function $\beta (k)$ reads 
\begin{eqnarray}
\label{betafunction}
\nonumber \beta (k) = \left(g_{0,1}^z(k) - g_{1,1}^r(k)\right)\Big[e^{-k(h+L_2 - L_1)} - e^{-k(h+L_2)}\\ - e^{-k(h-L_1)} + e^{-kh}\Big].
\end{eqnarray}
Here $\mu_0$ is the vacuum susceptibility and the g-functions are defined as $g_{\mu,i}^{\alpha}(k) = \int_0^{R_i}J_{\mu}(kr_i)r_im_{\alpha}(r_i)r_idr_i$, being $J_{\mu}(x)$ the Bessel function of first kind and order $\mu$. Note that the above equations are valid only for $h \geq L_1$. 

For the magnetic dot we used  magnetic parameters taken from the IrCoPt multilayers \cite{moreau2016}, i.e., a saturation magnetization  $M_s = 956$ kA/m, an anisotropy constant $K_u = 0.717$ MJ/m$^3$, a stiffness constant $A_\text{ex} = 10$ pJ/m and a DMI parameter $D = 1.6$ mJ/m$^2$ (for simplicity of notation, the sub-index 1 will be sometimes omitted to refer the physical magnitudes of the magnetic dot). Additionally, we  considered $R_1 = 100$ nm and $L_1 = 1$ nm,  that allows us to obtain a stable N\'eel skyrmion configuration \cite{tejo2018}. For the tip, and to produce an intense magnetostatic field, we choose a cobalt dot with a saturation magnetization $M_{s_2} = 1400$ kA/m, the same radius as the first dot, and a thickness $L_2 = 200$ nm. Hereafter, we will keep these geometrical and magnetic parameters fixed throughout the text. We will only vary the value of $R_2$ to obtain results shown in the Appendix \ref{E_R2}. 

The energy of the system composed by the magnetic tip and dot can be written as $E = E_\text{tip} + E_\text{dot} + E_\text{int}$, where $E_\text{tip}$ and $E_\text{dot}$ are the self-energies of the magnetic configuration of the tip and the dot, respectively. For the description of the skyrmion in the dot we use the trial function \cite{debonte1973} $m_z(r) =-P \cos \left[2\arctan\left( f(r)\right)\right]$, with $f(r) = (R_s/r)\exp\left(\frac{\xi}{l_{\text{ex}}}\left(R_s - r\right)\right)$, being $l_{\text{ex}} = \sqrt{2A_{ex}/\mu_0M_s^2}$, $\xi^2 = Q - 1$, with $Q = 2K_u/\mu_0M_s^2$ and $R_s$ the skyrmion radius, a minimizable parameter. The above trial function has been successfully used to describe such textures in previous works \cite{tejo2018,sheka2001} showing good agreement with micromagnetic simulations.

An important point is that by varying the skyrmion radius $R_s$, including values greater than $R_1$, it is possible to obtain the total energy for both the skyrmion and the homogeneous state considering the interaction with the magnetic tip. Indeed, for values of $R_s$ smaller than $R_1$ we have a skyrmion configuration, while the energy of a quasi uniform magnetic state with magnetization pointing along the $\pm z$-direction can be obtained for $R_s\geq R_1$.

Since we are interested in the effect of the magnetic tip on the magnetic configuration of the dot, the self-energy of the magnetic tip, whose magnetization is fixed during the process described below, only adds a constant value, so we can redefine the energy as $E = E_\text{dot} + E_\text{int}$. As already mentioned, to calculate $E_\text{dot}$ we use the skyrmion ansatz considering the exchange, dipolar, anisotropy and Dzyaloshinskii-Moriya contributions to the energy, which have already been calculated in Refs.\cite{vidal2017,tejo2018}, and $E_\text{int}$ is given by Eq.\eqref{Eint_main}. Through the calculations we will distinguish the orientation of the core ($P = \pm 1$) of the N\'eel skyrmion, its chirality ($C = \pm 1$) and also the magnetic direction of the tip. To check our results we have also performed micromagnetic simulations with the OOMMF software \cite {oommf}, for which we have considered cubic cells of $1\times 1\times 1$ nm$^3$ and the system has been prepared with two different magnetic configurations in the dot: a homogeneous state and a skyrmion state, from which we let the system relaxes to its global (or local) minimum energy state. This procedure was repeated by changing the distance between the dot and the tip, $h$, in steps of 1 nm.
\begin{figure}[h]
\centering
\includegraphics[width=0.5\textwidth]{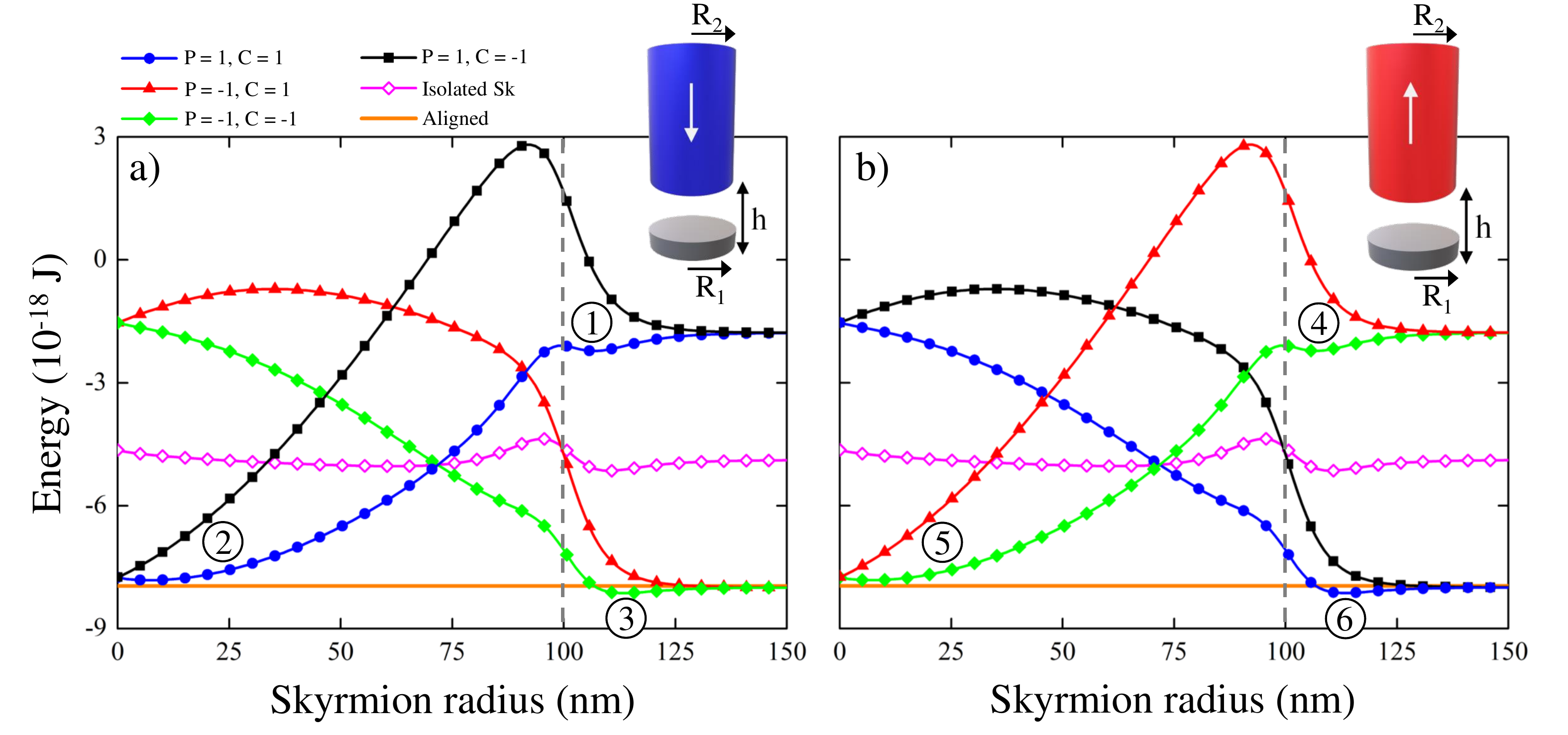}%
\caption{(Color online) Total energy for the different magnetic configurations in the dot when the magnetic tip is located at $h = 140$ nm distance with a homogeneous magnetization pointing along (a) $-\hat{z}$ and (b) $+\hat{z}$. The gray vertical dashed line represents the radius of the dot, $R_1$. In legend $P = \pm 1$ is referred to the skyrmion polarity and $C = \pm 1$ to its chirality. The numbers inside the circles denote specific situations that are explained in the text.}
\label{Fig1}
\end{figure}

We start our calculations obtaining the energy for different magnetic configurations in the dot. Fig.\ref{Fig1} illustrates  the energy, $E$, as a function of the skyrmion radius $R_s$ for $h=140$ nm when the magnetization of the tip points along a) $-\hat{z}$-direction and b) $+\hat{z}$-direction. In this figure the four possible combinations for the skyrmions polarity and chirality  are distinguishable, $C = \pm 1$ and $P = \pm 1$. The dashed vertical line denotes the limit of the dot ($R_1 = 100$ nm). It is important to notice that  results shown in Figs. 1a) and 1b) are equivalent  but with opposite chiralities and polarities. For example, in Fig.\ref{Fig1}a) the (blue on-line) dotted line shows the energy of a Ne\'el skyrmion hosted in the dot with a positive polarity and chirality, while in Fig.\ref{Fig1}b) the  energy, depicted by the solid diamonds line, exhibits the same behavior but considers  negative polarity and chirality.

Also in this figure we can notice that the interaction between the magnetic dot and the tip has a strong influence on the magnetic configurations that are allowed in the dot. For example, from the (pink on-line) hollow diamonds curve representing an isolated skyrmion, it is evidenced that when the dot interacts with the magnetic tip, for $h = 140$ nm, its energy changes according to the skyrmion chirality and polarity. Furthermore, since the lowest energy is depicted by the (green on-line) solid diamonds curve (point \circled{3}) or (blue on-line) dotted curve (point \circled{6}, the quasi-homogeneous state is the most  favorable energetically, having essentially the same energy than the full \textit{aligned} state in which the tip and dot magnetizations are parallel (nevertheless, see discussion below) . Also Fig. \ref{Fig1} shows that a skyrmion with a radius greater than the dot radius can be interpreted as a quasi-homogeneous state, which allow us to study simultaneously both configurations.

\begin{figure}
\centering
\vfill
\includegraphics[width=0.5\textwidth]{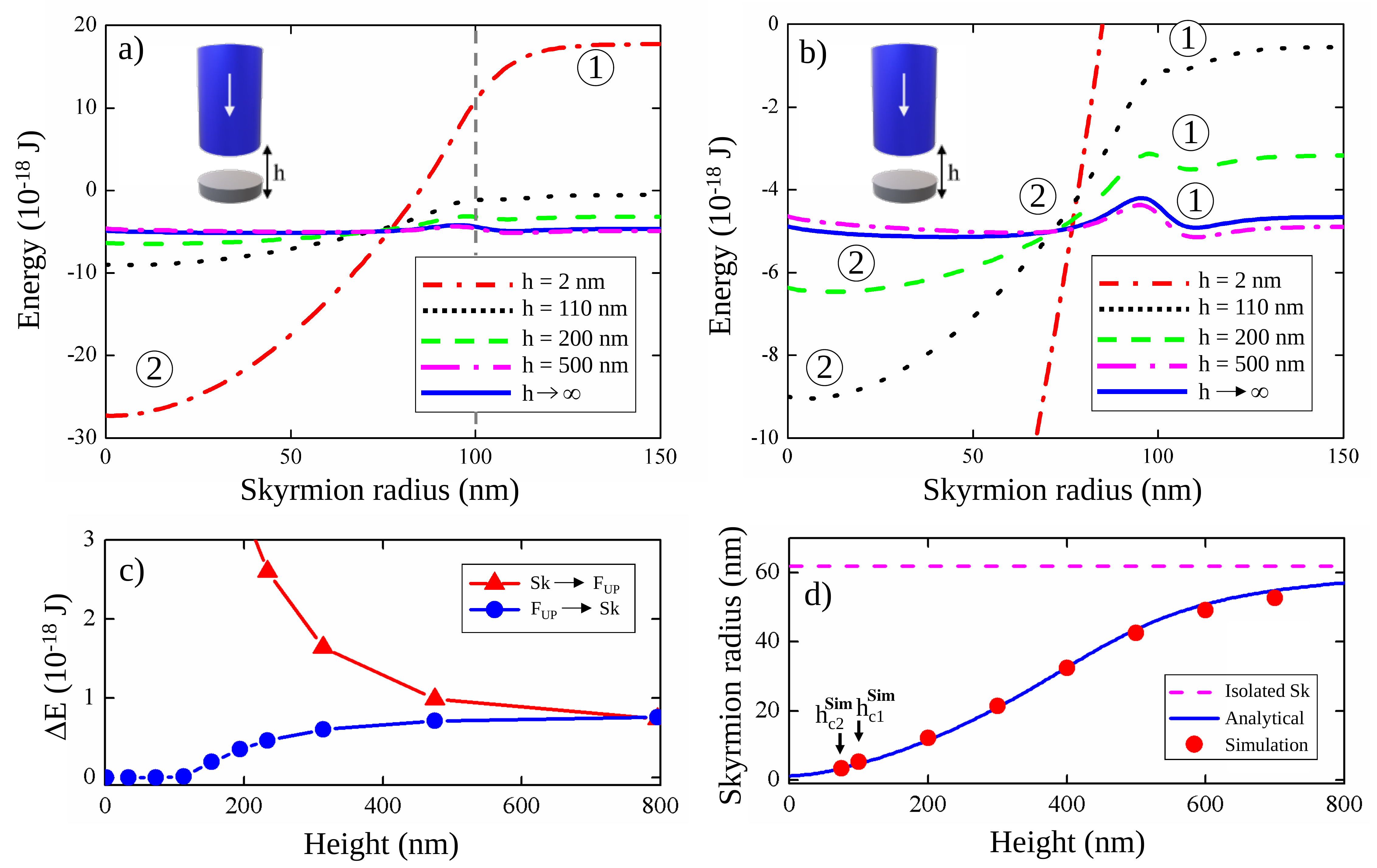}
\caption{(Color online) a) Skyrmion energy for $P = 1$ and $C = 1$ as function of the skyrmion radius for selected  heights, $h$. b) Zoom  of the  skyrmion energy  evidencing the presence (absence) of energy barriers for different heights. c) Energy barriers to go from a skyrmionic to a homogeneous state along $+\hat{z}$ (line with red on-line triangles) and from a homogeneous  along $+\hat{z}$ state to a skyrmionic configuration (line with blue on-line circles); and d) Skyrmion radius as a function of the vertical separation, $h$.  Red dots represent results from micromagnetic simulations. The heights $h_{c_2}^{sim}$ and $h_{c_1}^{sim}$ are explicitly depicted by arrows.}
\label{Fig2}
\end{figure}

In Fig.\ref{Fig2}a) we illustrate the skyrmion energy with $P = 1$ and $C = 1$ , which changes when varying the height $h$. As mentioned before, this configuration represents a quasi-homogeneous state pointing towards $+\hat{z}$ in the regime $R_s\geq R_1$. As expected,  for small values of $h$, the interaction energy becomes more relevant, substantially modifying the energy respect to the  the isolated or non-interacting case ($h\rightarrow \infty$). In fact, the skyrmion radius that minimizes the energy (point \circled{2}) changes when $h$ varies (see also Fig. \ref{Fig2}b)). 

To better understand the behavior of this system, we perform micromagnetic simulations, whose results are presented in Appendix \ref{Ap_reversal}, and find that the stability of the system depends not only on the vertical separation $h$, but also on the initial magnetic state of the system. That is, if initially the dot is in a quasi uniform magnetic state pointing along $+\hat{z}$ and then we approach to it (from very far away) a magnetic tip uniformly magnetized along $-\hat{z}$ then, depending on the vertical separation $h$, the dot could remain in the quasi uniform magnetization state (point $\circled{1}$ in Fig.\ref{Fig1}a)), change its magnetization to a skyrmionic texture with the core pointing along the opposite direction of the magnetization of the tip (point $\circled{2}$ in Fig.\ref{Fig1}a)), or reverses most of its magnetic moments until its magnetization is aligned with the magnetic moments of the tip (point $\circled{3}$ in Fig.\ref{Fig1}a)). Simulations also show that for $h> h_{c_1}^{sim} \approx 122$ nm, the dot remains with its magnetization opposite to the tip, while for $ h_{c_1}^{sim} > h > h_{c_2}^{sim} \approx 73$ nm, an isolated skyrmion with a small radius is nucleated with $P=1$ and $C=1$. Finally for $h<h_{c_2}^{sim}$ the magnetization of the dot is aligned with the tip magnetization. This behavior can be understood from the analytical results displayed in Fig.\ref{Fig2}b) which shows a zoom of the skyrmion energy presented in Fig. \ref{Fig2}a). In fact, in the absence of an external excitation to the system, besides the magnetic tip, the dot remains with a magnetization opposite to the tip unless the energy barrier from point $\circled{1}$ to point $\circled{2}$ vanishes, which occurs close to $h = h_{c_1}^{an}= 110$ nm, evidencing the good agreement between analytical and numerical results. For completeness, in Fig.\ref{Fig2}c) we show with a blue dotted line the behavior of this energy barrier as a function of the vertical separation. It is important to notice that once the skyrmion is stabilized, the energy cost to bring the dot back to the homogeneous state is considerably high, as shown by the line with red triangles in Fig. \ref{Fig2}c), manifesting the high stability of skyrmions.

Calculations also predict that the dot will evolve to a skyrmionic configuration when the energy barrier vanishes, even for values of $h$ smaller than $h_{c_2}^{sim}= 73$ nm (see red dashed-dotted and black dashed curves in Fig. \ref{Fig2}a)), while micromagnetic simulations show a magnetization reversal, as shown in Fig. \ref{Fig6} in the Appendix \ref{Ap_reversal}. To clarify this seemingly contradictory point, we display the size of the core of the skyrmion in Fig.\ref{Fig2}d) where the critical values $h_{c_1}^{sim}$ and $h_{c_2}^{sim}$ are explicitly shown with black arrows. Although according to analytical calculations for $h< h_{c_2}^{sim}$ the skyrmion radius is very small, namely $R_s \leq 3.6$ nm, if we compare it with the micromagnetic simulations, as we mentioned previously, we find that at these heights the dot does not stabilize a skyrmion. In fact, it reverts its magnetization from a quasi uniform state in $+\hat{z}$ to one in $-\hat{z}$ through the annihilation of a skyrmion by contraction of its core,  a behavior similar to that suggested by our analytical calculations, as evidenced in  Fig. \ref{Fig2}d). In this way our analytical model  predicts the behavior of the system for most of the values of  $h$ in the studied range,  and  exhibits some differences with simulations only for small $h$ values, that is, in the very interacting regime, where micromagnetic simulations predict the magnetization reversal of the dot. We attribute this difference to the mechanism by which the magnetization reversal  occurs, that is, possibly through the formation of a Bloch point as previously reported in Refs.\cite{lobanov2016,bessarab2018, siemens2016,verga2014,rohart2016}. A Bloch point is beyond the present formalism and, therefore, has not been considered here. However, the calculations show a clear tendency to a considerable decrease of  the skyrmion radius  when $h$ decreases, approaching a magnetization reversal.

The previous analysis is valid if we start with an isolated dot with a quasi uniform magnetization and we bring a magnetic tip with an antiparallel magnetization respect to the dot.  The equivalent  behavior illustrated in Figs.\ref{Fig1}a) and 1b) show that both the polarity and the chirality of the skyrmion nucleated in the dot can be handled by varying the direction of the magnetization of the tip and the dot. If the dot is magnetized along the $+\hat{z}$ ($-\hat{z}$) direction, approaching a tip with magnetization along the opposite direction will nucleate a skyrmion with $P = 1, C = 1$ ($P = -1, C = -1$). 
\begin{figure}
\centering
\includegraphics[width=0.33\textwidth]{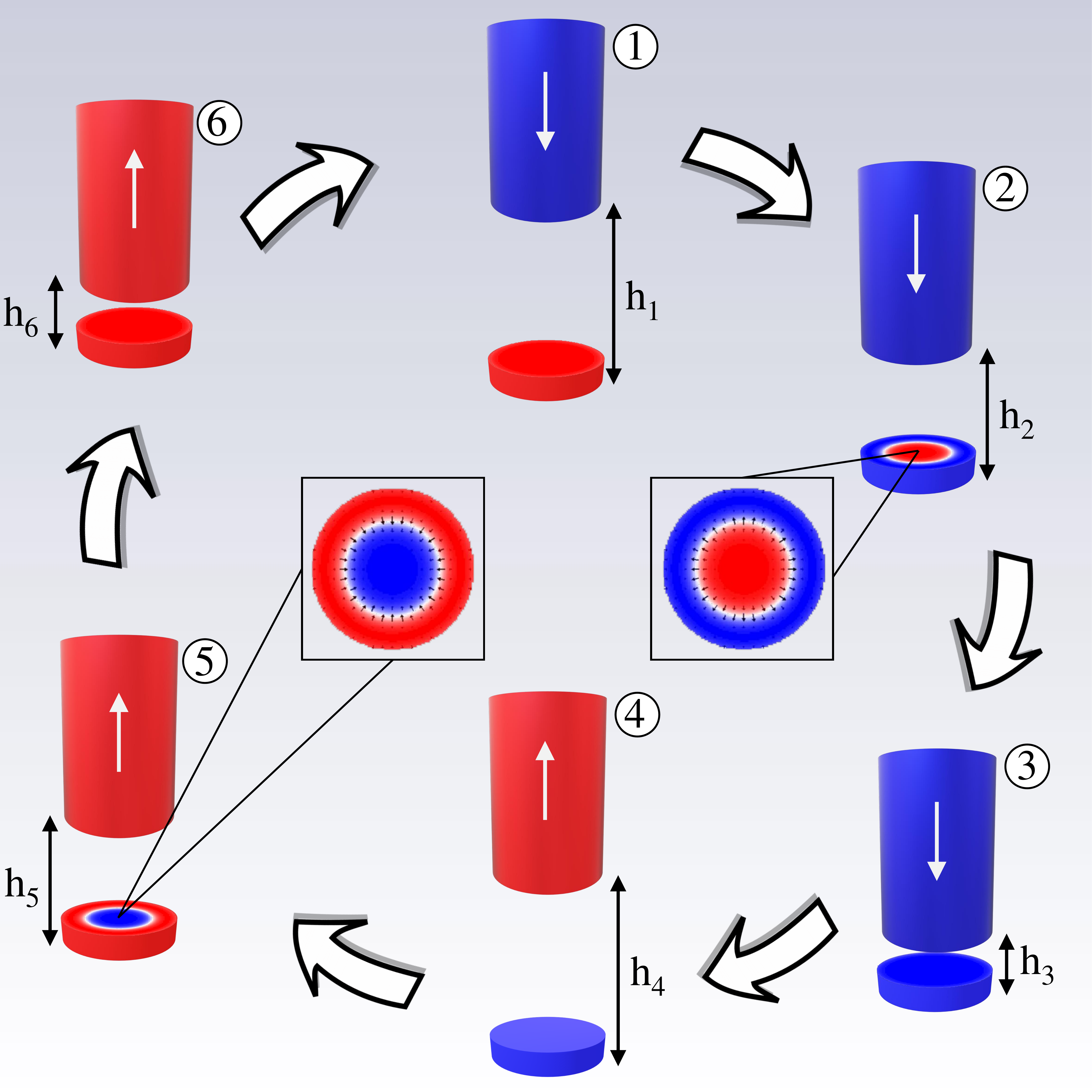}\hfill%
\caption{(Color online)  Proposed cycle to nucleate and annihilate a skyrmion in a dot with a specific polarity and chirality  by fixing the magnetization direction of the magnetic tip and its vertical separation with the dot.}
\label{Fig3}
\end{figure}

Now, if the dot has an isolated skyrmion as starting point, while approaching to it a magnetic tip magnetized along the opposite direction to the skyrmion core magnetization, the skyrmion will reduce its core size until all the magnetic moments (or most of them) are pointing in the same direction as the magnetic tip. Indeed, micromagnetic simulations show that the skyrmion radius decreases as the tip approaches the dot and the skyrmion remains stable until a critical value $h_{c_2}^{sim} = 72$ nm is reached. For values of $h$ smaller this critical value, the dot will exhibit an almost saturated magnetization parallel to the magnetization of the tip, reversing in the same way as explained above. The values of the skyrmion radius predicted by the simulations are illustrated in Fig.\ref{Fig2}d) by red points. It is important to mention that the simulations and analytical results are in very good agreement for the whole range $h > h_{c_2}^{sim}$. It is also interesting to note that if we choose a magnetic tip with a magnetization parallel to the core of the skyrmion then, as we approach the tip, the core size will increases, promoting the annihilation of the skyrmion through the core expansion. However, the range of values of $h$ for which the skyrmion is stable is smaller than the same situation when a magnetic tip is magnetized antiparallel to the skyrmion core direction.

\begin{figure}[h]
\centering
\includegraphics[width=0.25\textwidth]{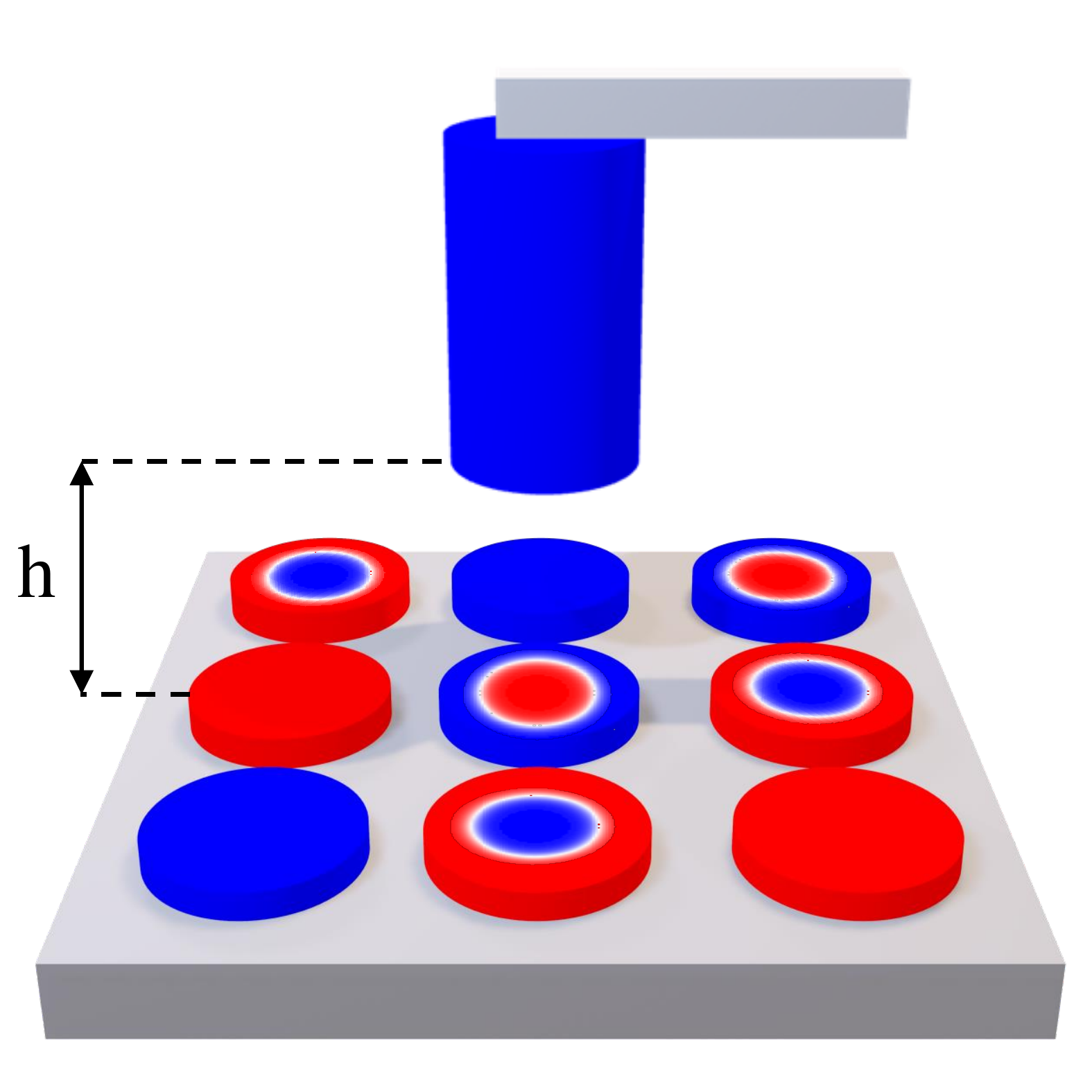}\hfill%
\caption{(Color online) Proposal for a skyrmion-based magnetic data storage device. Following the results explained throughout the text, this device could exploit the all degrees of freedom from the system allowing a $4^N$ storage capacity in terms of bits of information.}
\label{Fig4}
\end{figure}

The above results suggest a method, summarized in Fig.\ref{Fig3}, to nucleate/annihilate a skyrmion controlling its polarity and chirality.  For example, by starting at  $\circled{1}$ (the same  $\circled{1}$ shown in Fig.\ref{Fig1}) and approaching the magnetic tip to the dot, a skyrmion with positive polarity and chirality will ne nucleated ($\circled{2}$). As the magnetic tip approaches the dot, the latter will align its magnetization with the magnetization of the tip ( $\circled{3}$). We can go from  $\circled{3}$ to  $\circled{4}$ by removing the magnetic tip, inverting its magnetization, and then approaching again the tip to the dot. Then we can nucleate a skyrmion, but with negative polarity and chirality, as illustrated in  $\circled{5}$. If we continue approaching the magnetic tip to the dot, its magnetization will be parallel to the  magnetization of the tip ( $\circled{6}$). Removing the tip again, changing its magnetization direction, and approaching again  the dot, the cycle will be complete. This simple procedure can control all characteristics of a magnetic texture in the dot and could be the basis of a novel skyrmion-based data storage system. In fact, Fig.\ref{Fig4} illustrates how could operate such a device. If the magnetic tip is free to move vertically and horizontally, it can control the magnetization state of any dot in an array. The main feature of our proposal is that there are four stable states on each nanodot, which suppose four bits per each one in terms of data storage. Thus, for a $N\times N$ non-interacting nanodots array, the system could storage $4^N$ bits of highly stable information.

In conclusion, by means of analytical calculations and micromagnetic simulations we studied the effect that a cylindrical magnetic tip  on a magnetic dot. The inclusion of DMI in the dot allows to create skyrmionic textures in it. The magnetostatic field created by the tip can be used to control the polarity, chirality and size of the core of the skyrmion, or creating an almost homogeneous magnetic state, which is a particular case of a skyrmion with a core radius larger that the radius of the dot.  The different states can be achieved by varying the distance between the tip and the dot.   Our analytical results show a good agreement with micromagnetic simulations. In this way, and  from energy calculations, we can explain the dynamic associated to changes in the magnetostatic interaction. Our results allow us to propose a simple computer-based data storage device whose operating principle would be the control of the distance between the tip and the nanodot, and the magnetization of the tip.

We acknowledge support from Fondecyt under grants 1160198, 3180470, 1150952, 3190264 and Financiamiento Basal para Centros Cient\'ificos de Excelencia, CEDENNA, grant FB0807.


\appendix 

\section{General expression for the magnetostatic interaction energy between two cylindrical nanostructures with arbitrary magnetization}
\label{E_int_ap}
The interaction energy between two cylindrical nanostructures each one with radius $R_i$, thickness $L_i$ and magnetization direction $\mathbf{m}_i(\mathbf{r}_i) = m_{r,i}(\mathbf{r}_i)\hat{\mathbf{r}} + m_{z,i}(\mathbf{r}_i)\hat{\mathbf{z}}$, with $i= 1,2$,  separated a distance $s = \sqrt{d^2 + h^2}$ is \cite{escrig2008}
\begin{eqnarray}
\label{Eint_ap}
\nonumber\frac{E_{int}}{\mu_0M_{s1}M_{s2}} = -\frac{1}{2}\int_{0}^{2\pi}d\phi_2\int_{0}^{R_2}m_{r,2}(r_2)r_2dr_2\int_0^{\infty}dk\alpha(k)\times \\
\nonumber J_1\left(k\sqrt{r_2^2+d^2-2 r_2 d \cos\phi_2}\right)+\\
\pi\int_0^{\infty}dk g_{0,2}^z(k)J_0(k d)\beta(k),\hspace{1.3cm}
\end{eqnarray}
where the g-functions $g_{\mu,i}^{\alpha}(k)$ are defined in the main text and depend explicitly on the magnetic texture, while $\alpha(k)$ and $\beta(k)$ are functions defined as
\begin{widetext}
\begin{align}
\alpha(k) = \left\{ \begin{array}{cc} 
                \left(g_{0,1}^z(k) - g_{1,1}^r(k)\right)(-1+e^{kL_2})(-1+e^{kL_1})e^{-k(h+L_2)} & h\geq L_1, \\\\
                g_{0,1}^z(k)\Big[e^{-k(h+L_2)}(-1 + e^{kL_2})(-1 + e^{k(2h-L_1+L_2)})\Big] +\\ g_{1,1}^r(k)\Big[(-1+e^{kL_2})(e^{-k(L_1 - h)} + e^{-k(h + L_2)})- 2kL_2\Big]& h<L_1 \text{and} \\ & h+L_2\leq L_1\\
                g_{0,1}^z(k)\Big[2 - e^{-kh} - e^{-k(L_1 - h)} + e^{-k(h+L_2)} - e^{-k(h+L_2 - L_1)}\Big] + \\
 g_{1,1}^r(k)\Big[e^{-kh} - e^{-k(h-L_1)} + 2k(h-L_1)
 -e^{-k(h+L_2)} + e^{-k(h - L_1 + L_2)}\Big] & h<L_1 \text{and}\\ & h+L_2>L_1. 
                \end{array} \right.
\end{align}
and
\begin{align}
\beta(k) = \left\{ \begin{array}{cc} 
                \left(g_{0,1}^z(k) - g_{1,1}^r(k)\right)\Big[e^{-k(h+L_2 - L_1)} - e^{-k(h+L_2)} - e^{-k(h-L_1)} + e^{-kh}\Big] & h\geq L_1, \\\\
                g_{0,1}^z(k)\Big[e^{-k(L_1-L_2 - h)} - e^{-k(h+L_2)}  - e^{-k(L_1-h)} + e^{-kh}\Big] +   \\
g_{1,1}^r(k)\Big[e^{-k(L_1-L_2 - h)} + e^{-k(h+L_2)}  - e^{-k(L_1-h)} - e^{-kh}\Big]& h<L_1 \text{and} \\& h+L_2\leq L_1, \\
                g_{0,1}^z(k)\Big[-e^{-k(L_1 - h)} + e^{-k(h+L_2 - L_1)} + e^{-kh} - e^{-k(h+L_2)}\Big] +\\ 
 g_{1,1}^r(k)\Big[2-e^{-k(L_1 - h)} - e^{-k(h+L_2 - L_1)} - e^{-kh} + e^{-k(h+L_2)} \Big] & h<L_1 \text{and}\\ & h+L_2>L_1. 
               \end{array} \right.
\end{align}
\end{widetext}
\section{Skyrmion radius as a function of $h$ for different values of $R_2$}\label{E_R2}
Here we show how robust is our model when varying the radius of the tip.
\begin{figure}
\centering
\vfill
\includegraphics[width=0.45\textwidth]{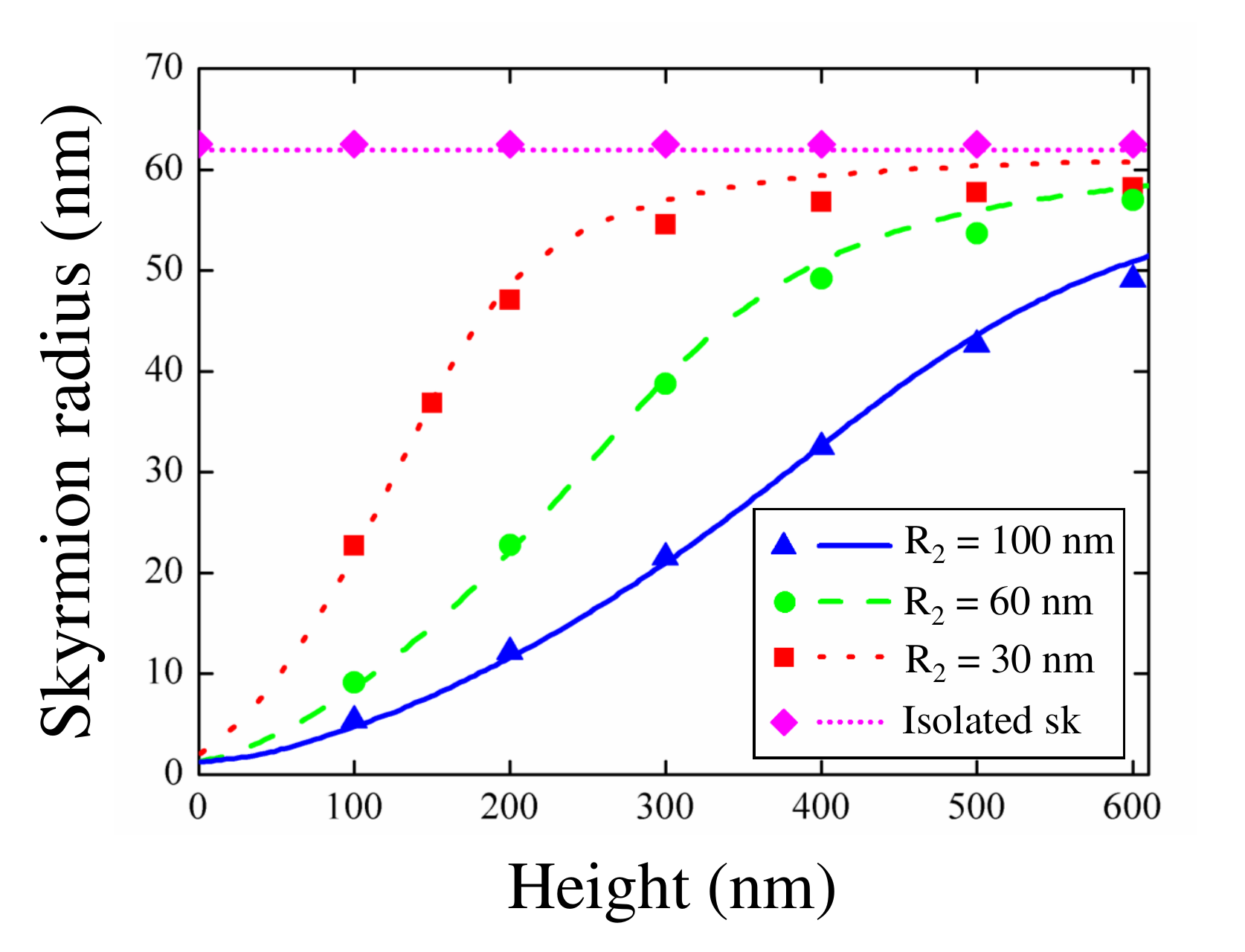}\hfill%
\caption{Skyrmion radius as function of the height, $h$,  for different values of $R_2$. Lines illustrate results from the analytical calculations while symbols represent results from micromagnetic simulations.}{\label{Fig5}}
\end{figure}
Fig. \ref{Fig5} illustrates the skyrmion radius as a function of the height $h$. It can be seen that results from both micromagnetic simulations and  analytical model are in good agreement  for the isolated and the interacting system. This allow us to conclude  that our model is also valid for different $R_2$ values, and that the general procedure proposed for a recording device, as well as the main features of the present formalism is independent in $R_2$. Mainly differences coming from the use  of other materials and radius will be given on $h^{sim}_c$ and $h^{an}_{c}$ values.

\section{Nanodot's magnetization reversal process induced by the magnetic tip} \label{Ap_reversal}

In this section and by means of micromagnetic simulations, we show the magnetization reversal process of a homogeneously magnetized nanodot when interacting with a magnetic tip. In particular, we show the case when the magnetization of the nanodot is pointing towards $+\hat{z}$, while the magnetization of the tip is antiparallel. 

\begin{figure}[H]
\centering
\vfill
\includegraphics[width=0.45\textwidth]{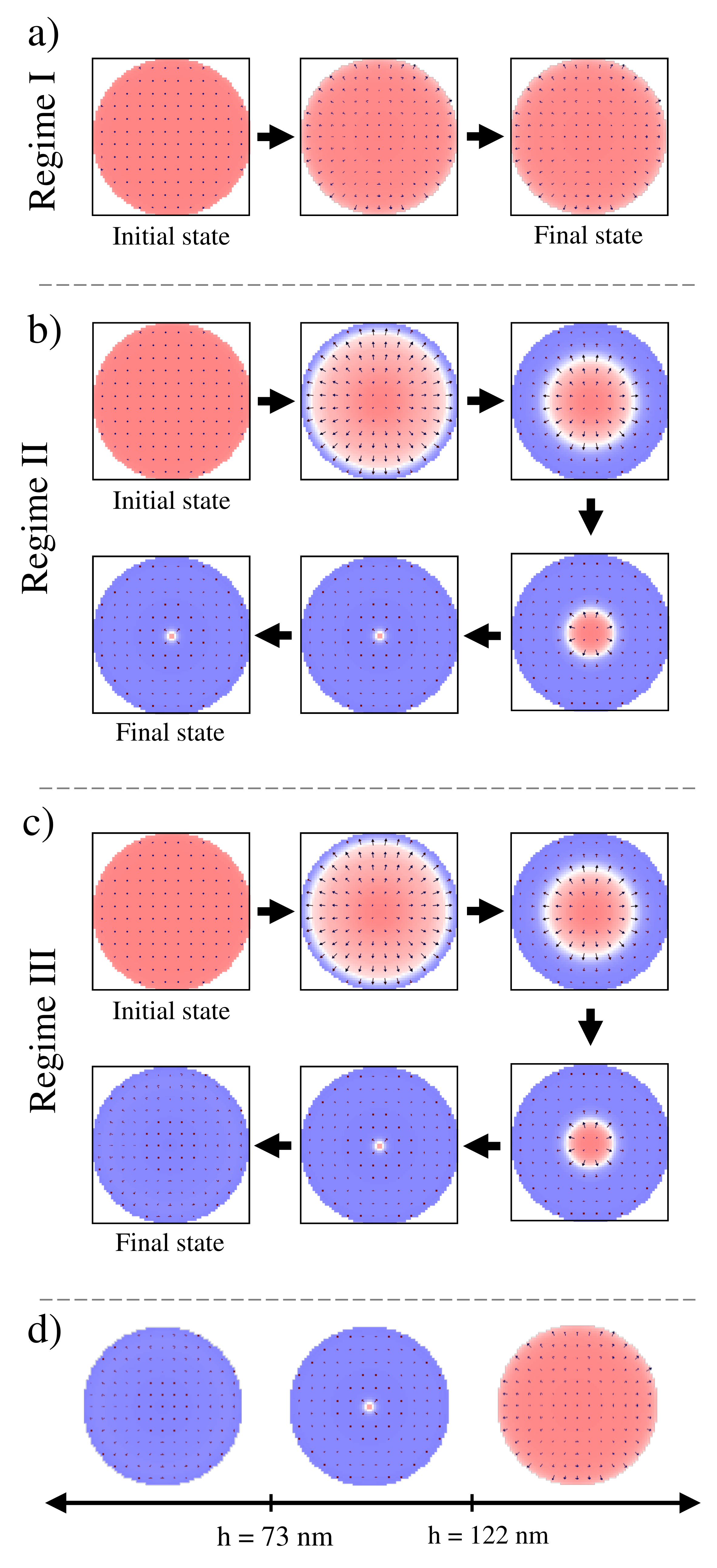}\hfill%
\caption{Magnetization reversal process of a magnetic nanodot magnetized along the $+\hat{z}$-direction interacting with a magnetic tip magnetized along the $-\hat{z}$-direction when the height of separation is a)$h = 125$ nm, b) $h = 80$ nm, and c) $h = 50$ nm. d) Summary of the three different regimes observed.}{\label{Fig6}}
\end{figure}

Fig. \ref{Fig6} shows the magnetization reversal process for three different heights $h$. Fig. \ref{Fig6}a) shows that for $h = 125$nm the magnetostatic field is not enough to revert the dot magnetization so the initial state is maintained. By decreasing the height $h$, Fig. \ref{Fig6}b) shows that for $h=80$ nm the initial state is no longer stable and a skyrmion is nucleated in the dot. In this case the magnetostatic field is not enough to completely inverts the dot magnetization and a skyrmion state is nucleated. If we continue decreasing $h$, the magnetization reversal process is fully completed, as shown in Fig. \ref{Fig6}c) for $h = 50$ nm. Simulations were carried out by varying the height $h$ in steps of 1 nm, allowing  to obtain an accurate range of $h$ at which the different magnetic states are stabie. Thus we found that, for these specific initial conditions, by approaching the magnetic tip until $h = h_{c1}^{sim}\approx 122$ nm, the system will remain in its initial state. If we decrease $h$ until  $h_{c_1}^{sim} > h > h_{c_2}^{sim} \approx 73$ nm, the system  evolves towards a skyrmion configuration with a very small radius, and for values $h<_{c_2}^{sim}$ the dot inverts its magnetization. These regimes are summarized in Fig. \ref{Fig6}d).

\begin{thebibliography}{99}
\bibitem{NatNanoF13} A. Fert, V. Cros and J. Sampaio, Nat. Nanotechn. \textbf{8}, 152 (2013).
\bibitem{zhang2015} X. Zhang, G. Zhao, H. Fangohr, J. Liu, W. Xia, and F. Morvan, Sci. Rep. \textbf{5}, 7643 (2015).

\bibitem{tomasello2015} R. Tomasello, E. Martinez, R. Zivieri, L. Torres, M. Carpentieri, and g. Finocchio, Sci. Rep. \textbf{4}, 6784 (2014).

\bibitem{NatNanoS13} J. Sampaio, V. Cros, A. Fert, S. Rohart and A. Thiaville, Nat. Nanotechnol. \textbf{8}, 839 (2013).

\bibitem{SciRep15} X. Zhang, M. Ezawa and Y. Zhou, Sci. Rep. \textbf{5}, 9400 (2015).

\bibitem{Iwasaki2013} J. Iwasaki, M. Mochizuki, and N. Nagaosa, Nat. Nanotechnol. \textbf{8}, 742 (2013).

\bibitem{Yu2012} X. Yu, N. Kanazawa, W. Zhang, T. Nagai, T. Hara, K. Kimoto, Y. Matsui, Y. Onose, and Y. Tokura, Nat. Commun \textbf{3}, 988 (2012).


\bibitem{SciRep16} W. Kang, Y. Huang, C. Zheng, W. Lv, N. Lei, Y. Zhang, X. Zhang, Y. Zhou and W. Zhao, Sci. Rep. \textbf{6}, 23164 (2016).

\bibitem{NJPhys17} Jan M{\"u}ller, New J. Phys. \textbf{19}, 025002 (2017).

\bibitem{JMMM94} A. Bogdanov and A. Hubert, J. Magn. Magn. Mater. \textbf{138}, 255 (1994).

\bibitem{NatMater15} I. K\'ezsm\'arki, S. Bord\'acs, P. Milde,E. Neuber, L. M. Eng, J. S. White, H. M. Ronnow, C. D. Dewhurst, M. Mochizuki, K. Yanai, H. Nakamura, D. Ehlers, V. Tsurkan and A. Loidl, Nat. Mater. \textbf{14}, 1116 (2015).

\bibitem{juge2018} R. Juge, S. Je, D. de Souza, S. Pizzini, L. Buda-Prejbeanu, L. Aballe, M. Foerster, A. Locatelli, and others, J. Magn. Magn. Mater. \textbf{455}, 3 (2017).

\bibitem{castro2016} M. Castro and S. Allende, J. Magn. Magn. Mater. \textbf{417}, 344 (2016)

\bibitem{boulle2016} O. Boulle, J. Vogel, H. Yang, S. Pizzini, D. de Souza, A. Locatelli, T. Mentes, A. Sala, L. Buda-Prejbeanu, and others, Nat. Nanotechnol \textbf{12}, 449 (2016).

\bibitem{guslienko2018} K. Guslienko, Appl. Phys. Express. \textbf{11}, 063007 (2018).

\bibitem{rohart2013} S. Rohart and A. Thiaville, Phys. Rev. B \textbf{88}, 184422 (2013).

\bibitem{DA} S. Castillo-Sep\'ulveda, R. M. Corona, A. S. Nu\~nez, and D. Altbir, J. Magn. Magn. Mater.  \textbf{443}, 116 (2019).

\bibitem{vidal2017} N. Vidal-Silva, A. Riveros, and J. Escrig, J. Magn. Magn. Mater. \textbf{443}, 116 (2017).

\bibitem{aharoni2000} A. Aharoni, \textit{Introduction to the Theory of Ferromagnetism}, vol. 109, Clarendon Press (2000).

\bibitem{riveros2018} A. Riveros, N. Vidal-Silva, F. Tejo, and J. Escrig, J. Magn. Magn. Mater. \textbf{460}, 292 (2018)

\bibitem{elias2017} R. G. El\'ias, N. Vidal-Silva, and A. Manchon, Phys. Rev. B, \textbf{95}, 104406 (2017).

\bibitem{tejo2018} F. Tejo, A. Riveros, J. Escrig, KY. Guslienko, and O. Chubykalo-Fesenko, Sci. Rep. \textbf{8}, 6280 (2018).

\bibitem{moreau2016} C. Moreau-Luchaire, C Moutafis, N. Reyren, J. Sampaio, C. Vaz, N. Van Horne, K. Bouzehouane, and others, Nat. Nanotechnol \textbf{11}, 444 (2016).

\bibitem{rooming2013} N. Romming, C. Hanneken, M. Menzel, J. Bickel, B. Wolter, K. von Bergmann, A. Kubetzka, and others, Science \textbf{341}, 636 (2013).

\bibitem{legrand2017} W. Legrand, D. Maccariello, N. Reyren, K. Garcia, C. Moutafis, C. Moreau-Luchaire, S. Collinn, K. Bouzehouane, and ohters, Nano Lett. \textbf{17}, 2703 (2017).

\bibitem{jiang2015} W. Jiang, P. Upadhyaya, W. Zhang, G. Yu, M. Jungfleisch, F. Fradin, J. Pearson, and others, Science \textbf{349}, 283 (2015).

\bibitem{zhou2014} Y. Zhou and M. Ezawa, Nat. Commu. \textbf{5}, 4652 (2014).

\bibitem{hsu2017} P. Hsu, A. Kubetzka, A. Finco, N. Romming, K. von Bergmann, and R. Wiesendanger, Nat. Nanotechnol. \textbf{12}, 123 (2017).

\bibitem{debonte1973} W. DeBonte, J. Appl. Phys. \textbf{44}, 1793 (1973).

\bibitem{sheka2001} D. Sheka, B. Ivanov, and F. Mertens, Phys. Rev. B \textbf{64}, 024432 (2001).

\bibitem{zhang2018} S. Zhang, J. Zhang, Q. Zhang, C. Barton, V. Neu, and others, Appl. Phys. Lett. \textbf{112}, 132405 (2018).

\bibitem{garanin18} D. Garanin, D. Capic, S. Zhang, X. Zhang, and E. Chudnovsky, J. Appl. Phys. \textbf{124}, 113901 (2018).

\bibitem{oommf} M. Donahue, OOMMF user's guide version 1.0 (1999).

\bibitem{lobanov2016} I. Lobanov, H. J\'onsson, and V. Uzdin, Phys. Rev. B \textbf{94}, 174418 (2016).

\bibitem{bessarab2018} P. Bessarab, G. Müiller, I. Lobanov, F. Rybakov, N. Kiselev, and others, Sci. Rep. \textbf{8}, 3433 (2018).

\bibitem{siemens2016} A. Siemenes, Y. Zhang, J. Hagmeister, E. Vedmedenko, and R. Wiesendanger, New J. Phys. \textbf{18}, 045021 (2016).

\bibitem{verga2014} A. Verga, Phys. Rev. B \textbf{90}, 174428 (2014).

\bibitem{rohart2016} S. Rohart, J. Miltat, and A. Thiaville, Phys. Rev. B \textbf{93}, 214412 (2016).

\bibitem{escrig2008} J. Escrig, S. Allende, D. Altbir, and M. Bahiana, Appl. Phys. Lett. \textbf{93}, 023101 (2008).
\end{thebibliography}

\end{document}